\documentclass{article}

\usepackage{arxiv}

\usepackage[utf8]{inputenc} 
\usepackage[T1]{fontenc}    
\usepackage{hyperref}       
\usepackage{url}            
\usepackage{booktabs}       
\usepackage{amsfonts}       
\usepackage{eqnarray,amsmath} 
\usepackage{nicefrac}       
\usepackage{microtype}      
\usepackage{lipsum}		
\usepackage{graphicx}
\usepackage{float}
\usepackage{doi}
\usepackage{natbib}
\setcitestyle{authoryear,open={(},close={)}} 
\newcommand{\citeA}[1]{\citeauthor{#1}}
\newcommand*\samethanks[1][\value{footnote}]{\footnotemark[#1]}

\title{Teleconnection patterns of different El Ni\~no types revealed by climate network curvature}

\author{Felix M. Strnad 
    \thanks{These two authors contributed equally to the paper.}\\
    Cluster of Excellence Machine Learning: New Perspectives for Science\\
    Universität Tübingen \\
    72072 Tübingen, Germany
    \AND
    Jakob Schlör 
    \samethanks[1]\\
    Cluster of Excellence Machine Learning: New Perspectives for Science\\
    Universität Tübingen \\
    72072 Tübingen, Germany
    \AND
    Christian Fröhlich\\
    Cluster of Excellence Machine Learning: New Perspectives for Science\\
    Universität Tübingen \\
    72072 Tübingen, Germany
    \AND
    Bedartha Goswami\\
    Cluster of Excellence Machine Learning: New Perspectives for Science\\
    Universität Tübingen \\
    72072 Tübingen, Germany
}

\renewcommand{\shorttitle}{\textit{Teleconnection patterns of different El Ni\~no types revealed by climate network curvature}}

\hypersetup{
pdftitle={Teleconnection patterns of different El Ni\~no types revealed by climate network curvature},
pdfsubject={arXiv-climnet},
pdfauthor={Jakob Schoer},
pdfkeywords={},
}

\begin{document}
\maketitle

\begin{abstract}
The diversity of El Ni\~no events is commonly described by two distinct flavors, the Eastern Pacific (EP) and Central Pacific (CP) types. While the remote impacts, i.e. teleconnections, of EP and CP events have been studied for different regions individually, a global picture of their teleconnection patterns is still lacking.
Here, we use Forman-Ricci curvature applied on climate networks constructed from 2-meter air temperature data to distinguish regional links from teleconnections. Our results confirm that teleconnection patterns are strongly influenced by the El Ni\~no type. EP events have primarily tropical teleconnections whereas CP events involve tropical-extratropical connections, particularly in the Pacific. Moreover, the central Pacific region does not have many teleconnections, even during CP events. It is mainly the eastern Pacific that mediates the remote influences for both El Ni\~no types.
\end{abstract}


\section{Introduction}

The El Ni{\~{n}}o Southern Oscillation (ENSO) is the largest interannual variation in the global climate system. It is a dynamical atmospheric and oceanic phenomenon characterized by anomalously warm (El Ni\~no) or cold (La Ni\~na)  phases of sea surface temperatures (SST) in the equatorial Pacific. Both phases are known to impact earth's climate significantly on long spatial scales, typically referred to as teleconnections \citep{Trenberth1997}. 

The interrelation between the `type' of El Ni\~no and its impact has been investigated in many studies over the past two decades \citep{Capotondi2015, Timmermann2018, Capotondi2020}. The diversity of El Ni\~no events is currently thought to be characterized by two modes: The ``canonical'' or Eastern Pacific (EP) El Ni\~no with peak SST anomalies in the eastern equatorial Pacific, and the ``El Ni\~no Modoki'' or Central Pacific (CP) El Ni\~no with peak SST anomalies in the central equatorial Pacific. Although the impact of EP and CP El Ni\~nos on different parts of the climate system --- such as the Indian Ocean (IO), maritime continent, tropical Atlantic, and Northern America --- have been studied thoroughly (see \citeA{Okumura2019} and \citeA{Taschetto2020} for an overview), previous work has mainly focused on single teleconnections of the El Ni\~no types. In comparison, little is known about differences in global teleconnection patterns between EP and CP. In this study, we close this gap by presenting a machine learning approach based on the Ricci-curvature of climate networks which detects global teleconnection patterns of the El Ni\~no types and highlights their differences.

Climate networks \citep{Dijkstra2019} have gained increasing interest for the analysis of spatial dependencies of climatic variables through their ability to reduce data to relevant climatic patterns. 
They have been widely used in the analysis of ENSO, starting with \citeA{Tsonis2008}, who investigated the topology of El Ni{\~{n}}o and La  Ni{\~{n}}a networks of surface air temperature. \citeA{Yamasaki2008} studied the global impact of El Ni\~no on various geographical zones while \citeA{Donges2009b} and \citeA{Zhou2015} examined geographical long-range teleconnections of ENSO. 

Impacts of ENSO diversity have been studied by evolving climate network analyses \citep{Radebach2013, Kittel2021}. \citeA{Wiedermann2016} use climate networks to find a robust way to distinguish different types of El Ni\~nos and La Ni\~nas. Similarly, \citeA{Lu2020} use an analysis of climate networks of the Pacific Ocean to distinguish EP and CP events and estimate their expected impacts.

Ricci-curvature measures the deviation of a continuous space from being locally flat and has been recently generalized to discrete spaces like complex networks \citep{ollivier2009ricci, sreejith2016forman}. In networks, the Ricci-curvature of a link describes the deviation of its surrounding from a regular grid (each node is connected to its four neighboring nodes). 
Ricci-curvature highlights whether an edge connects nodes within a community or bridges communities and thereby helps to understand visualize the network structure intuitively, applied for example in the analysis of financial markets \citep{Sandhu2016}, gene expressions \citep{Sandhu2015, Pouryahya2018}, brain connectivity \citep{Farooq2019}, urban transportation \citep{Gao2019}, power grids \citep{Jonckheere2019}, and epidemiology \citep{deSouza2021}. We use it to reveal global teleconnection patterns and are thereby able to uncover structurally different El Ni\~no impacts.

\section{Data and Methods} \label{sec:methods}

\subsection{Data} \label{sec:data}
We use daily 2-meter air temperature (or surface air temperature, SAT) data for the years 1950--2020 from the ERA5 Global Reanalysis database \citep{Hersbach2020}. We first detrend each time series, then subtract the daily climatology over the whole time period. We use next-neighbor interpolation to map the data to a grid of spatially approximately uniformly distributed points using the Fekete algorithm \citep{Bendito2007} to avoid spurious correlation patterns close to the poles \citep{Ebert-Uphoff2012}. The distance between grid points in the Fekete grid corresponds to 2.5° for points at the equator of a Gaussian grid, resulting in a total of $\approx 6000$ grid points.

\subsection{Classification of EP and CP El Ni\~no conditions}\label{sec:epcp}

We classify each day as EP (CP) based on whether the average Dec--Feb SST anomaly in the Ni\~no-3 region is greater (less) than that of the Ni\~no-4 region and Ni\~no-3 (Ni\~no-4) larger than 0.5 \citep{Capotondi2020}. Days that are neither EP, CP, nor La Ni\~na are labeled as `normal.' The SST anomalies are calculated with the Oceanic Ni\~no index baselines, i.e., using multiple centered 30-year base periods, successively updated in 5-year steps. For the end (beginning) of the time period, the last (first) available 30-year base period is used.

\subsection{2-meter air temperature Dec--Feb climate networks} \label{sec:cn}

Following earlier work on climate networks \citep{Donges2009, Wiedermann2016, Ciemer2020}, we estimate the weighted adjacency matrix $\mathbf{W}$ of the climate network by placing links between pairs of locations which have a correlation value among the 2\% strongest absolute correlations. Thus, for the correlation threshold $\rho_{0.98} = Q_{|\rho|} (0.98)$, where $Q_{X}(\cdot)$ denotes the quantile function for $X$, we define
\begin{eqnarray}
\mathbf{W}_{ij} = 
    \begin{cases}
    |\rho_{ij}|, & |\rho_{ij}| > \rho_{0.98} \\
    0, & \textrm{otherwise}
    \end{cases},
\end{eqnarray}
where $i,j \in \{1, \dots, N\}$ index spatial locations and $\rho_{ij}$ is the  Dec--Feb Spearman's rank-order correlation between locations $i$ and $j$. $\mathbf{W}_{ij}$ thus defines a network $G :=(V, E)$ with a set of edges (or links) $e_{ij} \in E$ that connect pairs of nodes $(v_i, v_j) \in V$ with link weight $w_{ij} = |\rho_{ij}|$.

Accounting for autocorrelation in the data, a two-sided test for non-random correlations at a confidence level of 1\% yields a threshold $|\hat{\rho}| = 0.35$. While our threshold is far higher, due to the high number of hypotheses tests ($3.6 \times 10^7$), we nevertheless expect a non-negligible number of network links to be false positives. We thus additionally use a spatial null model which assumes that correlations caused by physical mechanisms are likely part of spatially extended patterns \citep{Boers2019}. For each spatial location, we randomly rewire its network links $2000$ times and use a Gaussian kernel density estimator (KDE) to get the likelihood of a link to the chosen location. An observed network link to the chosen location is considered statistically significant if the spatial likelihood of the link (also obtained via Gaussian KDE) is above the $99.9$-th percentile of the local null model link distribution.

\subsection{Curvature of complex networks} \label{sec:ricci_curvature}
\begin{figure}[!tb]
    \centering
    \includegraphics[width=1.\linewidth]{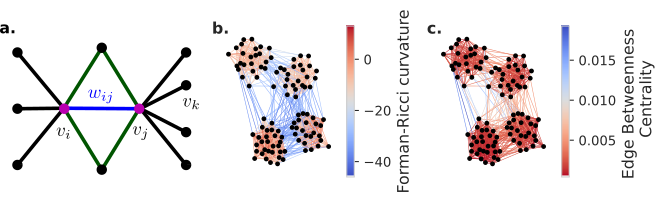}
    \caption{\textbf{Curvature of complex networks.} Forman-curvature of an edge with weight $w_{ij}$ connecting nodes $v_i$ and $v_j$ is obtained by Eq. \ref{eq:FormanCurvature} (\textbf{a}).     In the synthetic network constructed with a Stochastic Block Model, Forman-curvature (\textbf{b}) provides values to distinguish within-community links (red) from between-community links (blue) in contrast to betweenness centrality (\textbf{c}) with only a few links with high scores. The network with four communities are visualized using the spring layout from NetworkX \citep{NetworkX}.}
    \label{fig:toy}
\end{figure}
The Ricci-curvature of a network link describes how the connectivity of its network neighborhood differs from the connectivity of a regular grid. Out of two numerical approximations of Ricci curvature on networks --- Forman-Ricci curvature \citep{forman2003bochner, sreejith2016forman} and Ollivier-Ricci curvature \citep{ollivier2010survey} --- we use Forman-Ricci curvature (henceforth simply Forman curvature) as it is computationally cheaper and as both definitions are highly correlated, barring slight differences in extreme values \citep{Samal2018}. Forman curvature of edge $e_{ij}$ in an undirected network with weight $w_{ij} \in \mathbf{W}$ is estimated as,
\begin{eqnarray}
    \label{eq:FormanCurvature}
    F_{ij} = w_{ij} \left( 
        |\mathcal{T}_{ij}| \cdot w_{ij} 
        + \frac{2}{w_{ij}} 
        - \sum_{ \substack{k=1 \\ w_{ik} w_{kj}=0}}^{N} 
          \sum_{ \substack{l\in \{ i,j \} \\ w_{lk} > 0}} \frac{1}{\sqrt{w_{ij} \cdot w_{lk}}} 
        \right),
\end{eqnarray}
where $\mathcal{T}_{ij} := \{ v_k : w_{ik}w_{jk} > 0 \}$ denotes the set of nodes in the neighborhood of $v_i$ and $v_j$ which form triangles containing edge $e_{ij}$ (green edges in Fig.\,\ref{fig:toy}a) and $|\cdot|$ denotes set cardinality. The last term in Eq.\,\ref{eq:FormanCurvature} counts the number of edges adjacent to node $v_i$ and $v_j$ which do not form triangles with edge $e_{ij}$ (black edges in Fig.\,\ref{fig:toy}\,a). Equation \ref{eq:FormanCurvature} approximates the ``augmented'' Forman curvature (cf. \citeA{Samal2018}, Eq.\,9) by considering only triangles, no node weights, and no cycle weights. 

Since curvature is in general not symmetrically distributed around zero, we use the 10\% most positive curvatures,
\begin{eqnarray}
F_{ij}^+ &:= \{ F_{ij} : F_{ij} > Q_{F}(0.9) \},
\label{eq:extreme-curv}
\end{eqnarray}
and the 10\% most negative curvatures $F_{ij}^-$ (defined similarly as values less than $ Q_{F}(0.1)$. We also define the Forman-curvature of a network node,
\begin{eqnarray}
f_i := \frac{\sum_j F_{ij}}{\sum_j \mathbf{W}_{ij}}
\label{eq:node-curv}
\end{eqnarray}
to identify geographical locations, or `hotspots,' connected to strongly negatively or positively curved links. As the value ranges of curvature differs for different networks, we use the min-max transform to normalize it to $(-1, 1)$, denoted by $\tilde{F}_{ij}$. Corresponding to $\tilde{F}_{ij}^+$, we use Eq.\,\ref{eq:node-curv} to define $\tilde{f}_i^+$ as positive node-curvature hotspots and similarly, $\tilde{f}_i^-$ for negative node-curvature hotspots.

Consider a random network with four communities (Fig.\,\ref{fig:toy}\,b,\,c) generated using a stochastic block model from NetworkX \citep{NetworkX}. Forman-curvature clearly separates the between-community links from the within-community links. Within-community links are typically part of triangles, indicating local convergence of shortest paths, i.e. positive curvature. Conversely, links connecting nodes with a high degree that are not part of triangles indicate local divergence of shortest paths and negative curvature. Forman-curvature provides a continuous measure over network links that indicates if an edge is inside a community or if it straddles two communities. By comparison, the edge betweenness centrality (see SI sec \,\ref{sec:classical_network_measures})--- typically used to identify edges that connect communities --- fails to identify many between-community links (Fig.\,\ref{fig:toy}c). This is likely due to the binary notion of a shortest path - either a path is the shortest or it is not, implying that ``almost-shortest'' paths are not considered (SI Eq.\,\ref{eq:betweenness}).

\section{Results and Discussion} \label{sec:results}

\subsection{Spatial organization of teleconnections depends on El Ni\~no type} \label{sec:spatial_structure}
\begin{figure}[!b]
    \centering
    \includegraphics[width=1.\linewidth]{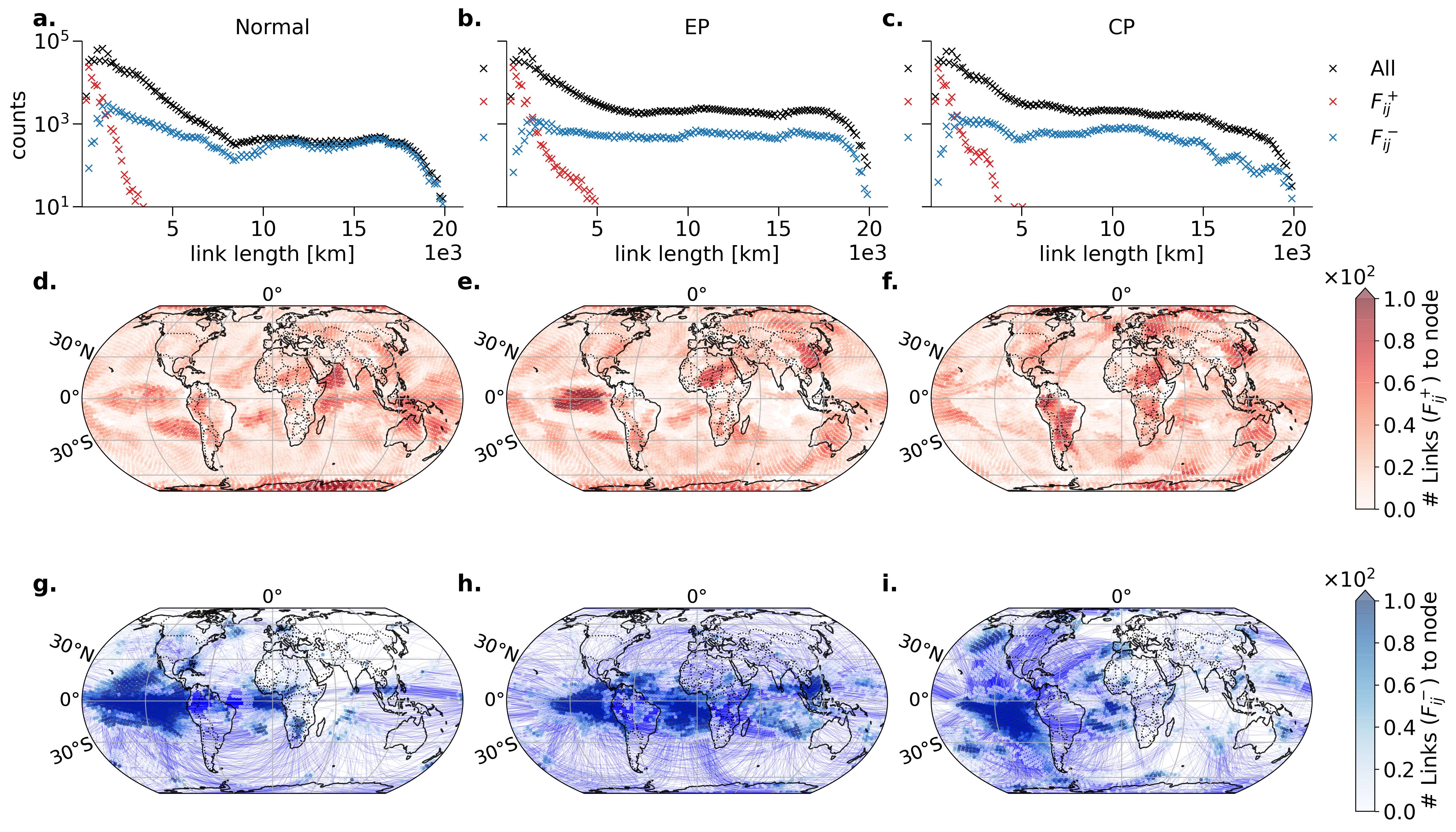}
    \caption{\textbf{Forman curvature of normal, EP, and CP El Ni\~no climate network links}. The networks are computed from 2m air temperature anomalies for normal (left column), EP(middle column) and CP (right column) conditions. The first row (\textbf{a}, \textbf{b}, \textbf{c}) depicts the spatial link length distribution for all (black), most positively $F_{ij}^+$ (red) and most negatively curved $F_{ij}^-$ (blue) edges. The second row (\textbf{d}, \textbf{e}, \textbf{f}) shows $F_{ij}^+$, the third row (\textbf{g}, \textbf{h}, \textbf{i}) $F_{ij}^-$. Colorbars indicate the number of edges to the node. For visual reasons only every 20th edge is plotted in \textbf{d}-\textbf{i}.
    }
	\label{fig:ll_distribution}
\end{figure}
Fig. \ref{fig:ll_distribution}\,a--c shows the distribution of great-circle lengths of the most positively curved $F_{ij}^+$ and most negatively curved $F_{ij}^- $ network links for normal, EP, and CP conditions respectively. While $F_{ij}^-$ links occur at all spatial scales, $F_{ij}^+$ links occur only at regional scales less than $5000$\,km. Long-range teleconnections are thus modulated almost always via negatively curved links; and as the curvature estimation of climate network links does not include any geographical information, likely, the correlation between negative climate network link curvature and long spatial scales is an intrinsic topological property of the SAT dynamics. Furthermore, the spatial distribution of $F_{ij}^+$  (Fig. \ref{fig:ll_distribution}\,d--f) and $F_{ij}^-$  (Fig. \ref{fig:ll_distribution}\,g--i) shows that while positive curvature occurs at only regional scales, negative curvature results in link bundles that are related to well-known teleconnection patterns, e.g., the connection between the eastern equatorial Pacific ENSO tongue pattern, and the tropical Atlantic.

The number of most negatively curved links for El Ni\~no conditions does not describe all spatially long-range links, in contrast to normal conditions (Fig. \ref{fig:ll_distribution}\,a--c). However, even with the lower number of links, $F_{ij}^-$ links undergo a drastic spatial reorganization for El Ni\~no conditions, especially from normal conditions (Fig. \ref{fig:ll_distribution}\,g) to CP conditions (Fig. \ref{fig:ll_distribution}\,i). During CP El Ni\~nos, e.g., the tropical Pacific shows an enhanced connection to the extratropical Pacific as well as to the mid-latitude North Atlantic region (see Sec.\,\ref{sec:enso_flavors}). Even for EP conditions (Fig. \ref{fig:ll_distribution}\,h), we observe that the connection between the tropical Pacific and the southern Atlantic is strengthened in comparison to normal months. Although we find subtle changes in the spatial organization of the most positively curved links between normal (Fig. \ref{fig:ll_distribution}\,d) and El Ni\~no conditions (Fig. \ref{fig:ll_distribution}\,e, f), such as weakening of regional correlation structures in the tropical Atlantic and strengthening of correlation in the West African monsoon belt, these changes are not as drastic as the ones that occur in the negatively curved links.

Note, we mainly discover teleconnections within the global oceans because correlations are generally higher over oceans than over land due to slower oceanic SAT variability \citep{Lambert2011}.

\subsection{EP El Ni\~no teleconnections are tropical, while CP El Ni\~no teleconnections are at all latitudes}
\begin{figure}[!tb]
    \centering
    \includegraphics[width=1.\textwidth]{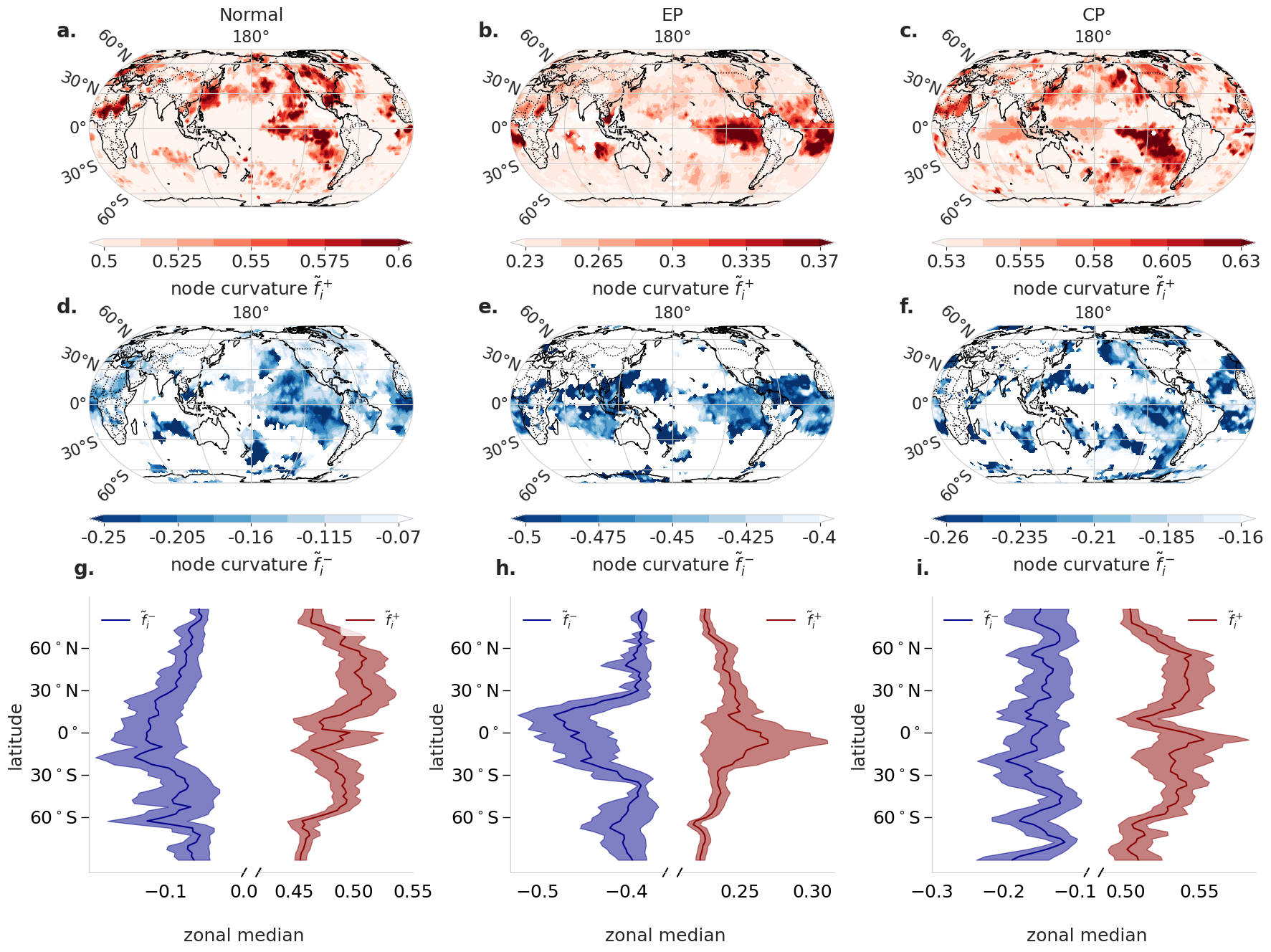}
    \caption{\textbf{Node curvature of normal, EP El Ni\~no, and CP El Ni\~no climate networks.} Node curvature for most positively curved, $\tilde{f}_i^+$ (\textbf{a-c}) and most negatively curved, $\tilde{f}_i^-$ (\textbf{d-f}) links are shown. EP hotspots (middle column) are more constrained to the tropics than for CP (right column). $\tilde{f}_i^{+/-}$ for normal conditions (left column) shows more similarity to CP than to EP. Zonal medians of $\tilde{f}_i^+$ (red) and $\tilde{f}_i^+$ (blue) further confirm this similarity. (\textbf{g-i}).}
    \label{fig:node-curvature}
\end{figure}
Positive curvature hotspots, $\tilde{f}_i^+$, for EP conditions reveal the well-known ENSO tongue (Fig.\,\ref{fig:node-curvature}b) typically observed in empirical orthogonal function analysis of sea surface temperature data \citep{Johnson2013}. We also find pronounced regions of $\tilde{f}_i^+$ in the IO and tropical Atlantic, which are known to be affected by strong EP El Ni\~no events \citep{Klein1999, Zhang2015, Rodrigues2015}. Under CP conditions, $\tilde{f}_i^+$ is spread over all latitudes and over different regions of the globe (Fig.\,\ref{fig:node-curvature}c). For instance, we observe a hotspot in the tropical Pacific similar to the El Ni\~no tongue, which is however shifted towards the dateline and also extended southwards. The spatial pattern under CP conditions is more similar to normal months (Fig.\,\ref{fig:node-curvature}a) than under EP conditions. $\tilde{f}_i^+$ shows strong similarity to node degree (SI \ref{fig:node_degree}).

Negative node-curvature hotspots, $\tilde{f}_i^-$, for EP conditions (Fig.\,\ref{fig:node-curvature}e) show enhanced teleconnections arising from South China Sea, tropical IO, eastern tropical Pacific, and the tropical Atlantic. This coincides with a decrease in teleconnections arising from the extratropical Pacific, southern IO, North Atlantic (near Greenland), and the Southern Ocean. As with $\tilde{f}_i^+$, here too the spatial pattern of CP conditions (Fig.\,\ref{fig:node-curvature}f) is spread all over the globe and looks more similar to normal conditions (Fig.\,\ref{fig:node-curvature}d), barring the northern tropical Pacific and the North Atlantic, both of which show a decrease in teleconnections.

Overall, EP conditions result in confinement of hotspots around the tropics for both regional links and teleconnections, which is further confirmed by comparing the zonal medians shown in Fig.\,\ref{fig:node-curvature}\,g--i. CP conditions, on the other hand, show only minor differences from normal months, although the magnitude of curvature values increase (SI \ref{fig:node_curvature_distribution}). This partly corroborates previous work \citep{Wiedermann2016,Lu2020} which also reports a strong localization of climate network links during EP conditions. These studies, however, did not focus on the impact of ENSO diversity on climate network teleconnections.

\subsection{EP and CP El Ni\~no teleconnection patterns of eastern and central Pacific Ocean, IO, and Labrador Sea} \label{sec:enso_flavors}
\begin{figure}[!tb]
    \centering
    \includegraphics[width=.8\textwidth]{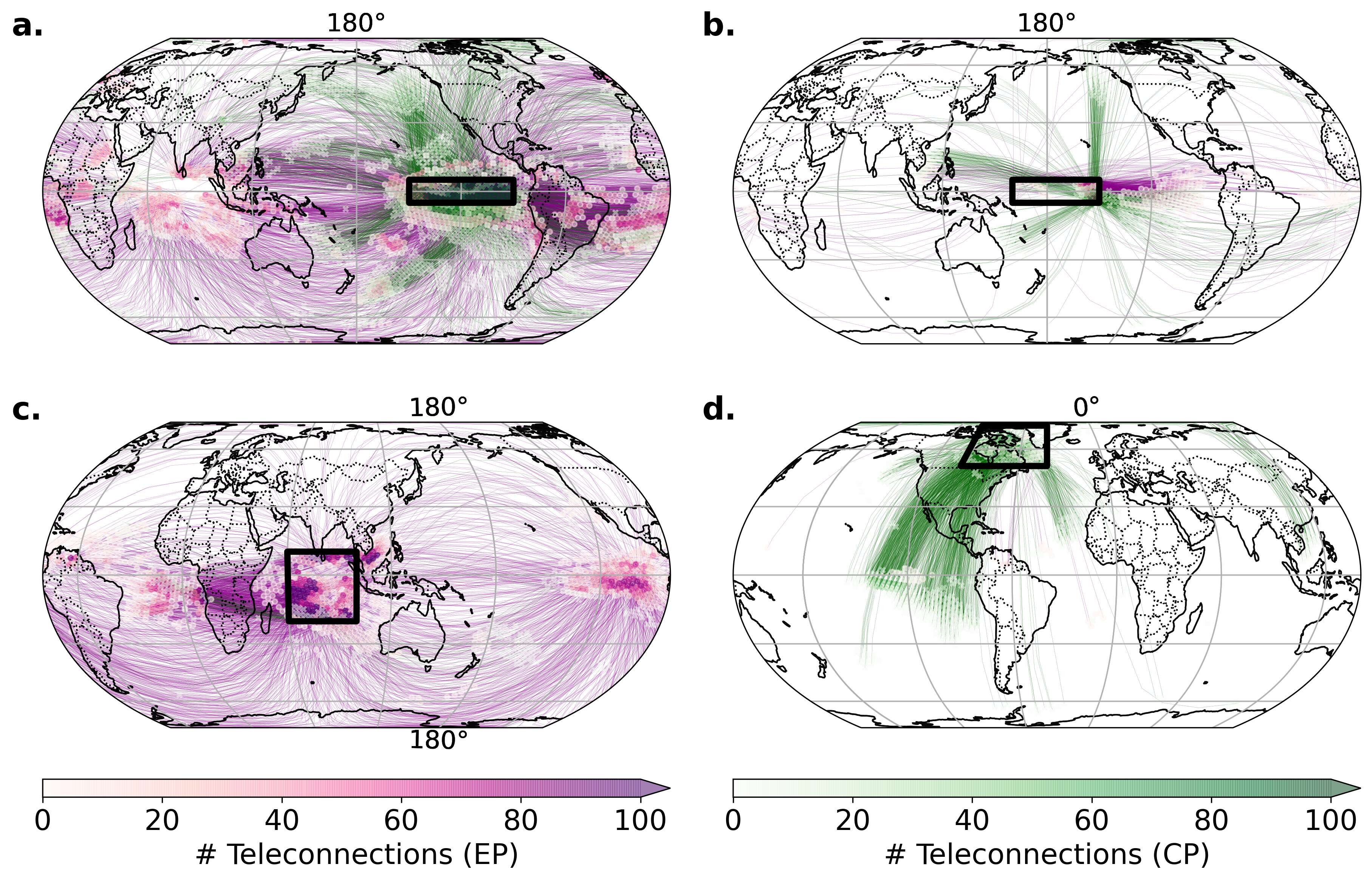}
    \caption{\textbf{Teleconnections arising from eastern and central Pacific Ocean, IO, and Labrador Sea.} 
    $F_{ij}^-$, i.e. teleconnections, for EP (purple) and CP (green) El Ni\~no events connected to the four selected regions (black rectangles): Ni\~no 3 (\textbf{a}), Ni\~no 4 (\textbf{b}), IO (\textbf{c}), and Labrador Sea (\textbf{d}). For visual clarity, only a third of all links are shown.}
    \label{fig:edges-regions}
\end{figure}
Teleconnections link the Ni\~no 3 region (Fig.\,\ref{fig:edges-regions}a) to the tropical Atlantic during both EP and CP conditions, supporting earlier work showing that strong El Ni\~nos can lead to warming in the tropical Atlantic mediated by the tropospheric temperature mechanism \citep{Chang2006} and the atmospheric bridge via the Pacific North Atlantic (PNA) pattern \citep{Rodrigues2011}. There are also links from the Ni\~no 3 region that appear only under EP conditions, such as those that connect to the IO, likely arising because of the influence of eastern tropical Pacific SSTs on the IO during and after El Ni\~no events \citep{Klein1999}. Some links from the Ni\~no 3 box arise only under CP conditions, primarily connecting the extratropical Pacific. These links are likely due to the North and South Pacific Meridional Mode (NPMM and SPMM), as atmospheric and oceanic anomalies in the extratropics associated with SPMM and NPMM affect the intensity and flavor of El Ni\~no \citep{You2018}. The relationship of NPMM and SPMM with ENSO diversity is, however, still under debate \citep{Amaya2019}.

The teleconnections of the Ni\~no 4 region (Fig.\,\ref{fig:edges-regions}b) are far fewer compared to the Ni\~no 3 region (Fig.\,\ref{fig:edges-regions}a). It is rather surprising that the Nin\~o 4 region is not well connected even during CP conditions, although a CP El Ni\~no is primarily characterised by higher SST anomalies in the Ni\~no 4 region than in the Ni\~no 3 region. A likely explanation is that although the temperature fields are more anomalous in the central Pacific during CP conditions, the impacts of these anomalies are nevertheless mediated via the eastern Pacific. We find the same result for climate networks created from networks with shorter time ranges as well. 

The IO has a large number of most negatively curved links during EP conditions but not for CP (Fig. \ref{fig:node-curvature}\,c,\,d). The EP event teleconnections link the IO to the tropical Atlantic and Pacific basin (Fig. \ref{fig:edges-regions}\,c). The links to the tropical Atlantic could be either attributed to indirect links mitigated by the impact of the Nin\~o 3 region to the tropical Atlantic or might resemble direct impact between the oceans as recently described by \citeA{Zhang2021}. CP conditions do not result in teleconnections in the IO as CP events are generally weaker \citep{Zhang2015}. 

The Labrador Sea (Fig.\,\ref{fig:edges-regions}d) is another pronounced area of most negative node-curvature in the CP network not present in the EP network (compare Fig. \ref{fig:node-curvature}\,c,d). Edges with the most negative curvature adjacent to the Labrador Sea connect to the eastern tropical Pacific and the extra topical Atlantic. This pattern may be attributed to the North Atlantic Oscillation (NAO), which refers to sea level pressure changes in the Arctic and subtropical Atlantic \citep{Jimenez-Esteve2018}. El Ni\~no is known to cause a negative NAO pattern driven by the PNA pattern where a negative NAO phase is attributable to CP El Ni\~no events via the subtropical bridge \citep{Graf2012, Domeisen2019}.

\begin{figure}[!tp]
    \centering
    \includegraphics[height=1.\textwidth]{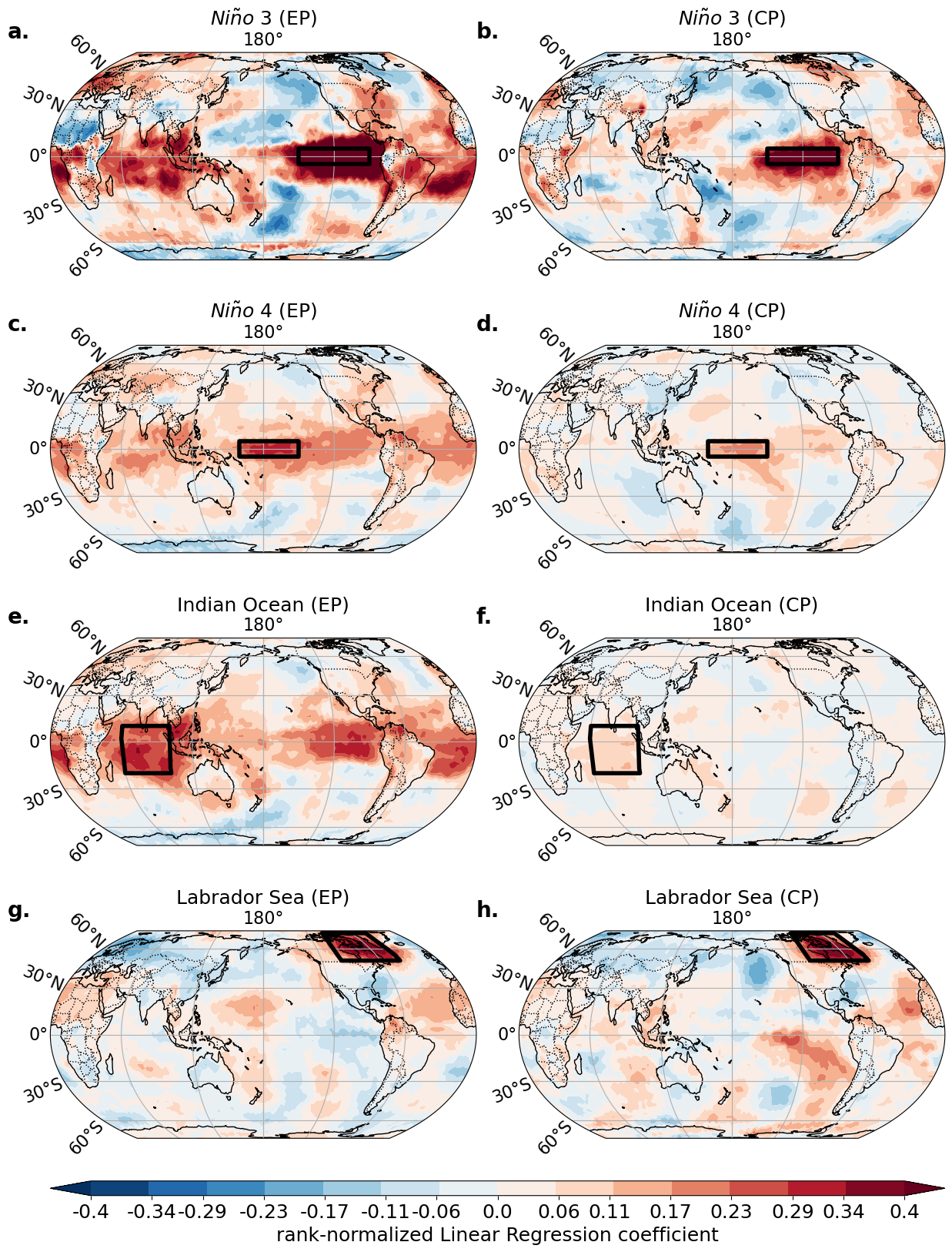}
    \caption{\textbf{SAT fields regressed on temperatures from the four selected regions.}
    Rank-normalized SAT time series at each location is regressed against all the rank-normalized SAT time series inside the four regions (black rectangles, same as in Fig.\,\ref{fig:edges-regions}) for EP events (left column) and CP events (right column) and the average linear regression coefficient is shown.}
    \label{fig:LR-coefficient}
\end{figure}

Regressing the SAT fields against the data in the four boxes
corroborates the findings from the curvature-based analysis (Fig.\,\ref{fig:LR-coefficient}, SI Fig.\,\ref{fig:LR-coefficient_notnorm}, \ref{fig:LR-coefficient_standardize}, \ref{fig:LR-coefficient_0_1_normalized}). The Ni\~no 3 region is highly teleconnected during both EP and CP conditions, but the spatial organization is different between the two (Fig.\,\ref{fig:LR-coefficient}\,a,\,b). The tropical Indian and Atlantic oceans are well modeled during EP and the tropical-extratropical links in the Pacific and the Labrador Sea are well modeled during CP. By contrast, the Ni\~no 4 box explains very little of the temperature fields irrespective of the El Ni\~no type (Fig.\,\ref{fig:LR-coefficient}\,c,\,d). Similarly, we find that the IO is highly active during EP but not during CP conditions (Fig.\,\ref{fig:LR-coefficient}\,e,\,f); and that the Labrador Sea is correlated to the tropics only during CP (Fig.\,\ref{fig:LR-coefficient}\,g,\,i). Note that we can to identify the teleconnection patterns only because we have the results from the curvature-based climate network analysis to guide our interpretations of the correlation maps. Without the curvature analysis, i.e. using simple correlation maps (SI Fig.\,\ref{fig:average_correlation}) or classical complex network measures (SI Fig.\,\ref{fig:node_degree}, \ref{fig:betweenness}, \ref{fig:clustering}), it is not trivial to figure out the most important regions for each El Ni\~no flavor.

\section{Conclusion} \label{sec:conclusion}

We presented a new approach to estimate global teleconnection patterns of 2-meter air temperature and used it to investigate the teleconnections of Eastern Pacific and Central Pacific El Ni\~no events. Our approach involves correlation-based climate networks and a recently established network measure based on Ricci-curvature. In particular, we used Forman-Ricci curvature to distinguish links related to small-scale regional structures (positively curved links) from long-range teleconnections connecting regions from different parts of the globe (negatively curved links).

We showed that El Ni\~nos diversity drastically impacts the spatial organization of teleconnections. Using node-based curvature, we identified teleconnection hotspots for both EP and CP conditions. Our results showed that teleconnections in the EP climate network were mainly confined to the tropics, whereas CP network teleconnections were found at northern and southern mid-latitudes as well. We further investigated the impact of ENSO diversity on the teleconnection patterns of four specific regions: the Ni\~no 3 region, the Ni\~no 4 region, the northern IO, and the Labrador Sea. We found that the Ni\~no 3 region in the eastern Pacific has a large number of teleconnections irrespective of whether we consider EP or CP conditions, whereas the Ni\~no 4 region in the central Pacific has, by comparison, very few teleconnections under both EP and CP conditions. We thus conclude that the eastern Pacific is the primary mediator of El Ni\~no impacts irrespective of the El Ni\~no type, and acknowledge that further work on the role of the eastern Pacific during CP El Ni\~nos is needed. We found that the northern IO and the Labrador Sea show teleconnections almost only under EP and CP conditions respectively. While the teleconnections of the IO region to the Ni\~no 3 region during EP conditions are well-known, the links between the Labrador Sea to the eastern tropical Pacific and northern tropical Atlantic are not fully understood.

\section*{Open Research}
Datasets for this research are available from Copernicus Climate Change Service. The data from 1979 till date was taken from \citeA{ERA5} and the data from 1950 to 1979 from \citeA{ERA5Extension}. The code for generating and analyzing the networks is made publicly available under \citeA{CodeClimnet}. The code for reproducing the analysis of the network curvature described in this paper is publicly available under \citeA{CodeCurvature}.  

\section*{Acknowledgment}
Funded by the Deutsche Forschungsgemeinschaft (DFG, German Research Foundation) under Germany’s Excellence Strategy – EXC number 2064/1 – Project number 390727645. The authors thank the International Max Planck Research School for Intelligent Systems (IMPRS-IS) for supporting Jakob Schlör and Felix Strnad.

\bibliography{./library.bib}

\newpage
\appendix
\section*{Supplementary Information}

\section{Complex Network Measures} \label{sec:classical_network_measures}
Assume complex network graph $G$, defined by its set of nodes $V$, connected by its set of edges $E$, where $e_{ij}$ denotes an edge connecting node $v_i$ to node $v_j$. This type of a graph is described by its adjacency matrix $\mathbf{A}$:
\begin{eqnarray*}
\mathbf{A}_{ij} = 
    \begin{cases}
    1, & e_{ij} \in E \\
    0, & \textrm{otherwise}
    \end{cases},
\end{eqnarray*}
If the edges $e_{ij}$ are weighted by weights $w_{ij}$, the weighted graph is described by
\begin{eqnarray*}
\mathbf{W}_{ij} = 
    \begin{cases}
    w_{ij}, & e_{ij} \in E \\
    0, & \textrm{otherwise}
    \end{cases},
\end{eqnarray*}

\paragraph{Node degree}
The node degree $\Tilde{k}_i$ and the weighted node degree $k_i$ is a frequently applied measure to study complex networks \citep{Donges2009b, Boers2013, Donges2015}. It counts the (weighted) number of neighboring nodes to node $v_i$, defined by: 
\begin{align}
    k_i &= \sum_{j}^N \mathbf{W}_{ij}  \; , \label{eq:node_degree} \\
    \Tilde{k}_i &= \sum_{j}^N \mathbf{A}_{ij}  \; ,
\end{align}
where $N$ denotes the number of nodes $v_i \in V$.

\begin{figure}[!tb]
    \centering
    \includegraphics[height=\textwidth]{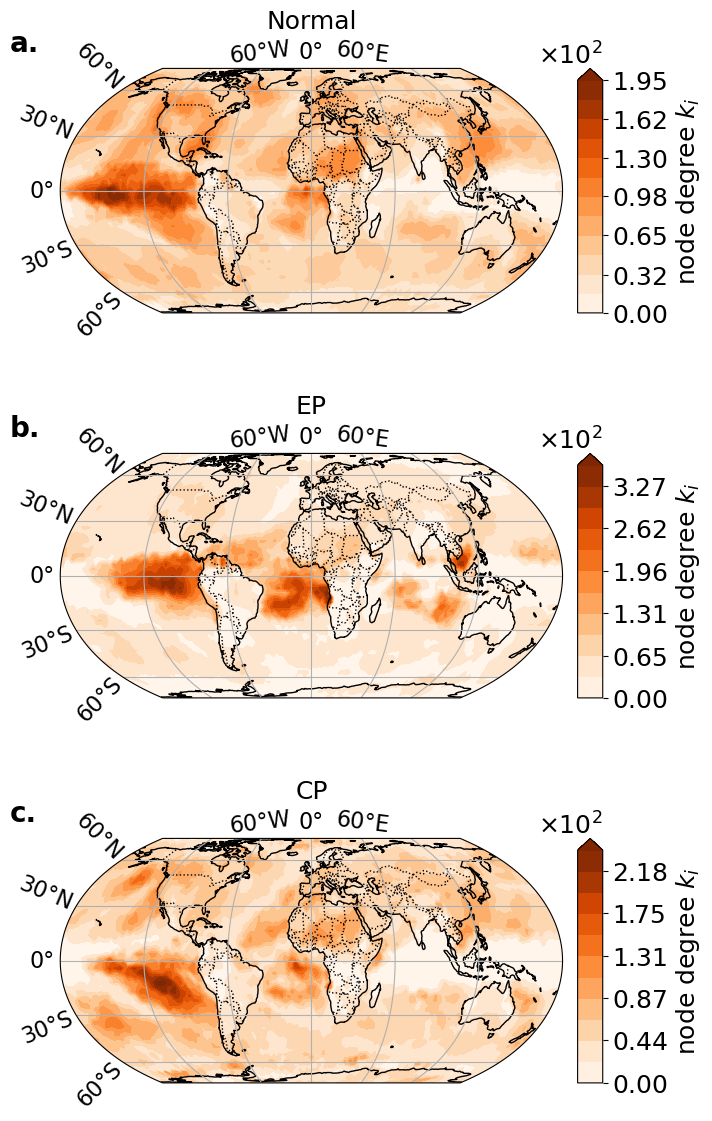}
    \caption{\textbf{Node degree for normal, EP and CP conditions}. The weighted node degree is computed following Eq.\, \ref{eq:node_degree}. The climate networks are obtained for normal (\textbf{a}), on Eastern Pacific El Ni\~no (\textbf{b}) and on Central Pacific El Ni\~no (\textbf{c}) conditions.}
    \label{fig:node_degree}
\end{figure}

In Figure \ref{fig:node_degree} the node degree is visualized for normal (\ref{fig:node_degree}\, \textbf{a}), EP (\ref{fig:node_degree}\, \textbf{b}) and CP (\ref{fig:node_degree}\, \textbf{c}) networks. 

\begin{figure}[!tb]
    \centering
    \includegraphics[height=1.\textwidth]{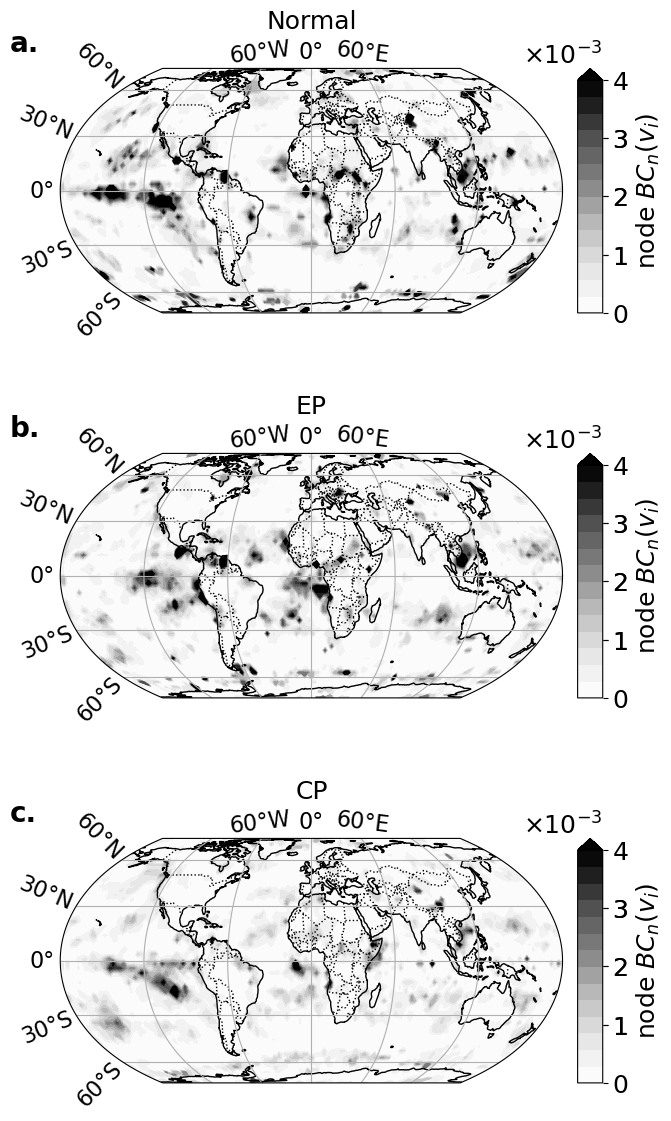}
    \caption{\textbf{Node Betweenness Centrality for normal, EP and CP conditions}. The betweenness centrality is computed using Eq.\, \ref{eq:node_betweenness}. The climate networks obtained for normal (\textbf{a}), on Eastern Pacific El Ni\~no (\textbf{b}) and on Central Pacific El Ni\~no (\textbf{c}) conditions.}
    \label{fig:betweenness}
\end{figure}

\paragraph{Betweenness Centrality}
Betweenness Centrality is defined for nodes $v$ and edges $e$ as,
\begin{align}
  BC_v(v_i) &= \sum_{s,t}^N \frac{\sigma(v_s, v_t|v_i)}{\sigma(v_s, v_t)} \; ,  \label{eq:node_betweenness} \\
  BC_e(e_{ij}) &= \sum_{s,t}^N \frac{\sigma(v_s, v_t|e_{ij})}{\sigma(v_s, v_t)} \; , \label{eq:betweenness}
\end{align}
where $\sigma (v_s,v_t)$ denotes the number of shortest paths between nodes $v_s$ and $v_t$ and $\sigma(v_s,v_t | v_i) \leq \sigma(v_s,v_t)$ the number of all shortest paths that include node $v_i$.  Similarly for edge betweenness centrality  $\sigma(v_s,v_t | e_{ij}) \leq \sigma(v_s,v_t)$ yields the number of all shortest paths that include edge $e_{ij}$. For instance, betweenness centrality measures were used in \citep{Freeman1977, Donges2009, Boers2013, Ciemer2018}. $BC_n$ is often referred to as pathway of a variable through the network and therefore taken as an indicator for the flow of the variable of interest. 
In Figure \ref{fig:betweenness} the node degree is visualized for normal year (\ref{fig:betweenness}\, \textbf{a}), EP (\ref{fig:betweenness}\, \textbf{b}) and CP (\ref{fig:betweenness}\, \textbf{c}) conditions.

\paragraph{Clustering Coefficient}
The clustering coefficient of a node $v_i$ is defined by the fraction of possible triangles through the actual number of nodes in the network normalized by the unweighted node degree $\Tilde{k}_i$:
\begin{align}
   c_i = \frac{2\mathcal{T}(v_i)}{\Tilde{k}_i(\Tilde{k}_i-1)} \label{eq:clustering} \; .
\end{align}
Here, $\mathcal{T}(v_i)$ describes the number of triangles including node $v_i$. 

\begin{figure}[!tb]
    \centering
    \includegraphics[height=1.\textwidth]{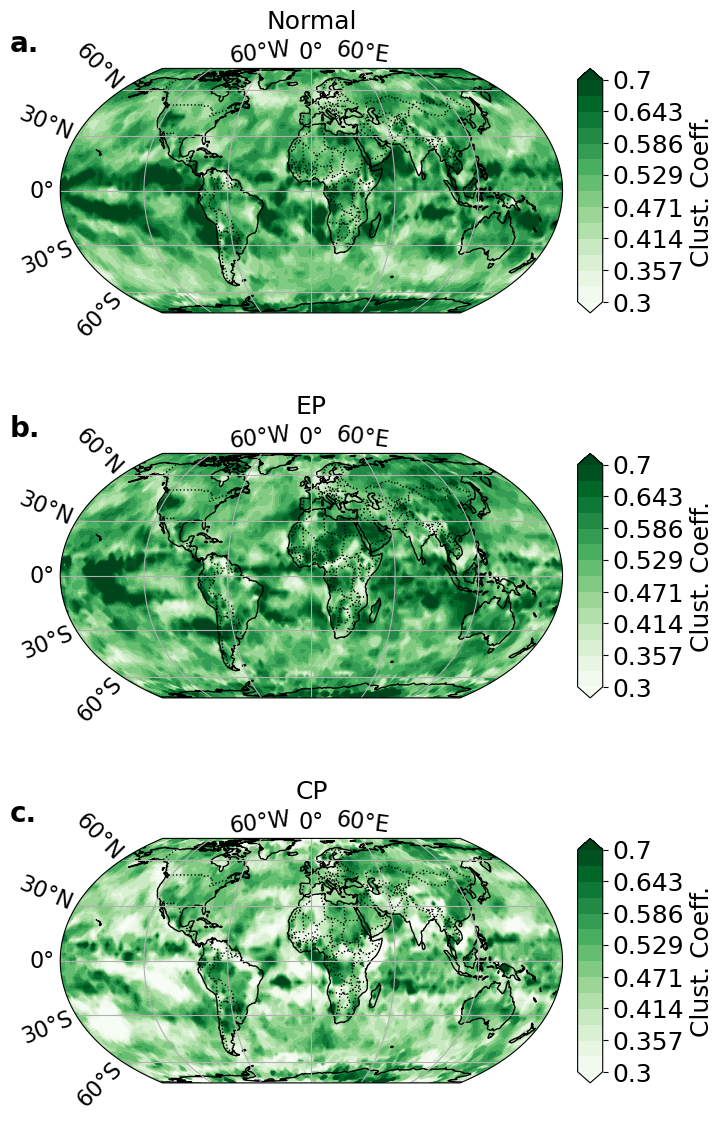}
    \caption{\textbf{Clustering Coefficient for normal, EP and CP conditions}. The clustering coefficient is computed using Eq.\, \ref{eq:clustering}. Networks are obtained for normal (\textbf{a}), EP (\textbf{b}) and CP (\textbf{c}) conditions.}
    \label{fig:clustering}
\end{figure}

\begin{figure}[!tb]
    \centering
    \includegraphics[width=1.\textwidth]{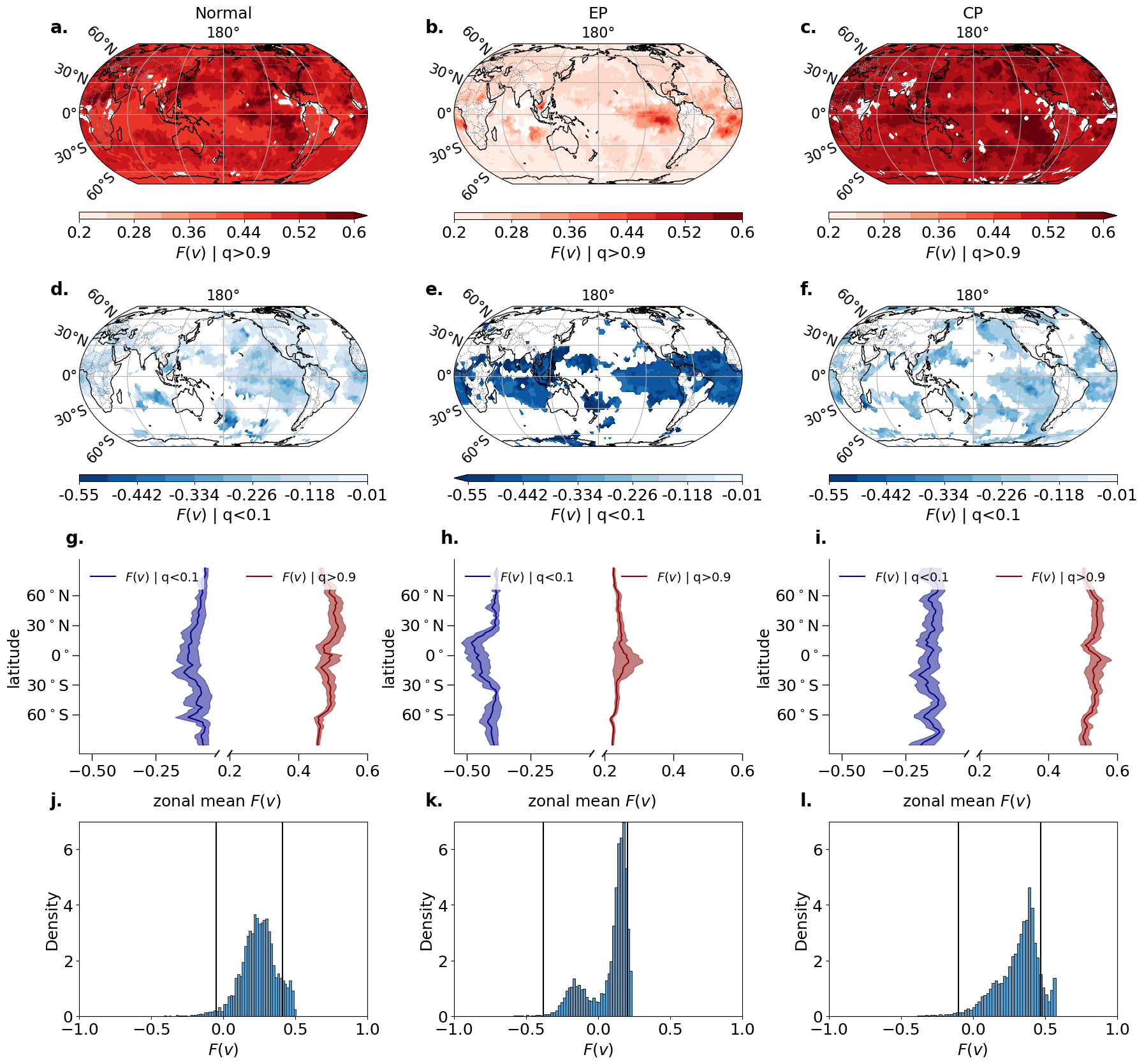}
    \caption{\textbf{Same as Fig. 3 but with equal scales}. The network curvature scales are fixed between the networks \textbf{(a-i)}. The differences between the distributions of node curvature $\Tilde{f}_i$ of normal \textbf{(j)}, EP \textbf{(k)} and CP \textbf{(l)} networks could be attributed to their network topologies. The quantiles $Q_F(0.1)$ and $Q_F(0.1)$ are indicated by black lines \textbf{(j-l)}. In order to plot the quantiles of node curvature we have to adapt the scales, as done in the main text Fig. 3.}
    \label{fig:node_curvature_distribution}
\end{figure}

\begin{figure}[!tb]
    \centering
    \includegraphics[height=1.\textwidth]{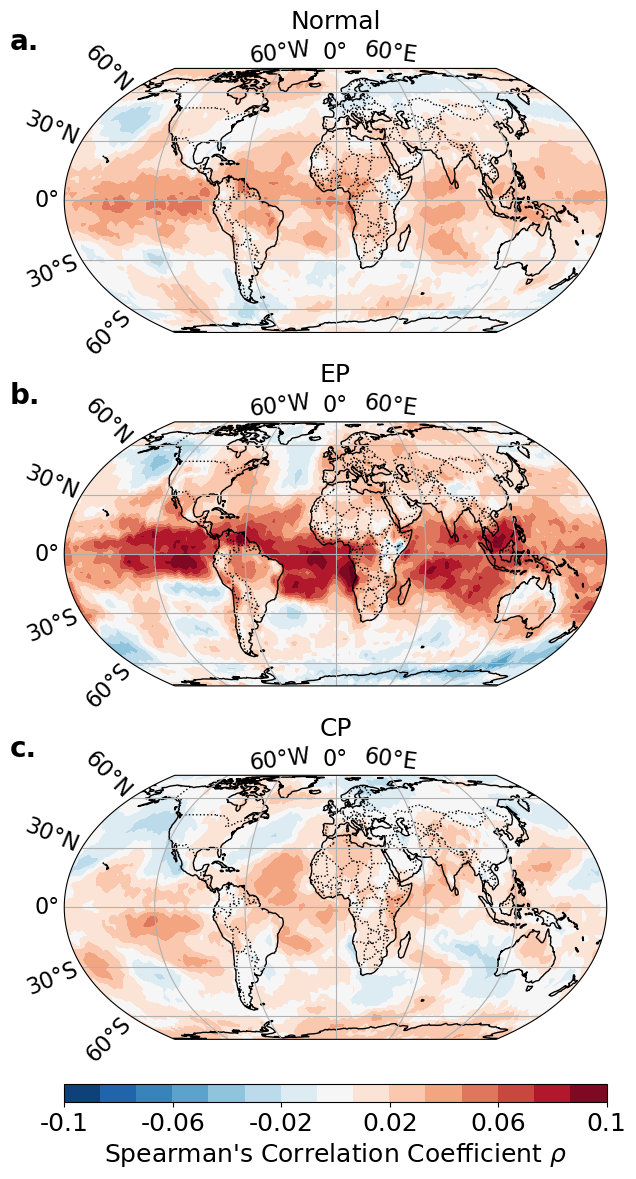}
    \caption{\textbf{Average Spearmann's correlation coefficient}. SAT time series at one location is Spearman-correlated against all other time series of the global dataset. The average Spearman's correlation coefficient is shown for the climate networks constructed based on normal (\textbf{a}), Eastern Pacific El Ni\~no (\textbf{b}) and Central Pacific El Ni\~no (\textbf{c}) conditions.}
    \label{fig:average_correlation}
\end{figure}

\begin{figure}[!tb]
    \centering
    \includegraphics[height=1.\textwidth]{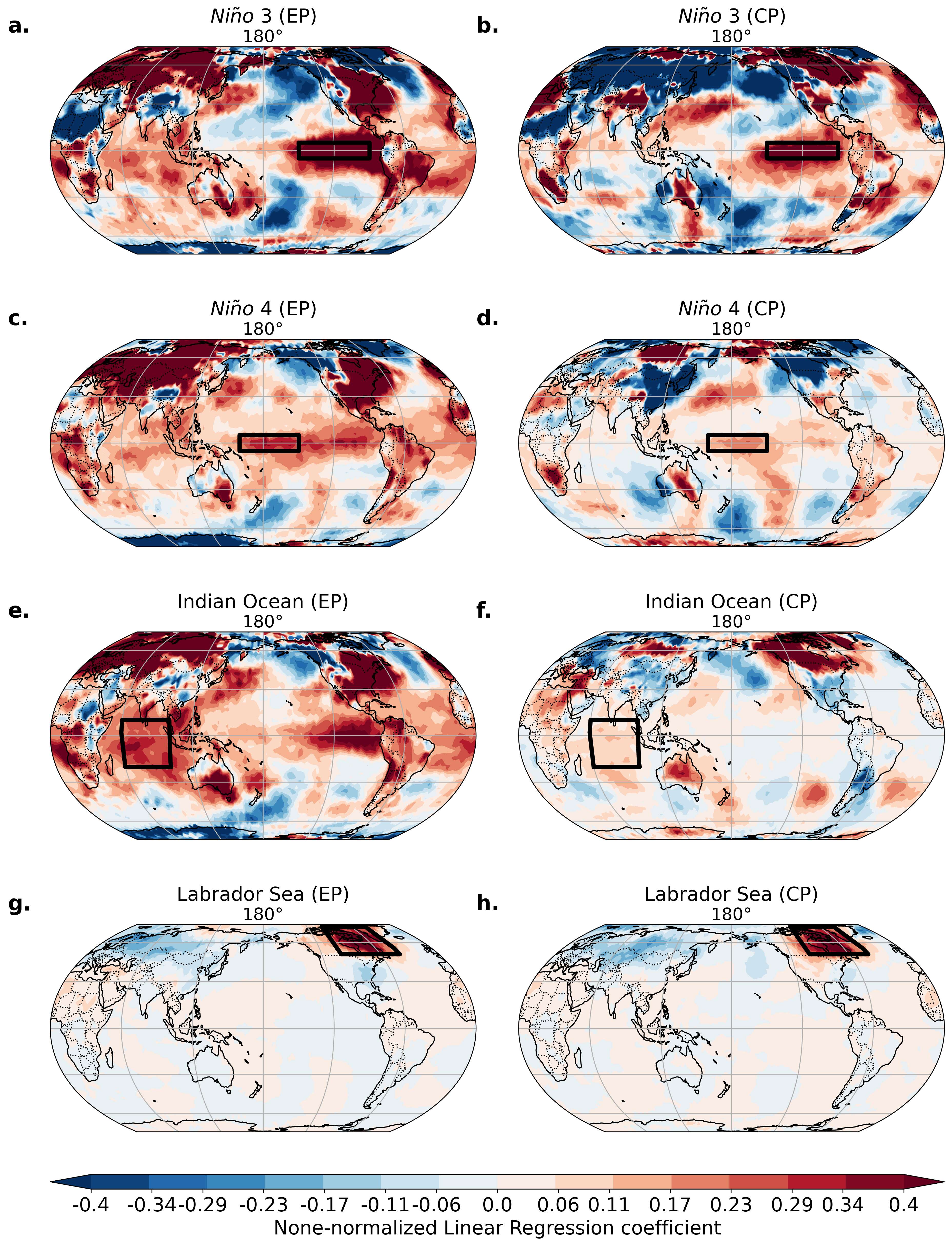}
    \caption{\textbf{SAT fields regressed on temperatures from the four selected regions.}
    SAT time series (not normalized) at each location is regressed against all the rank-normalized SAT time series inside the four regions (black rectangles). Displayed are the Ni\~no 3 box (1. row), Ni\~no 4 box (2. row), Indian Ocean (3. row), and Labrador Sea (4. row), with first column denoting EP and second column CP conditions.}
    \label{fig:LR-coefficient_notnorm}
\end{figure}

\begin{figure}[!tb]
    \centering
    \includegraphics[height=1.\textwidth]{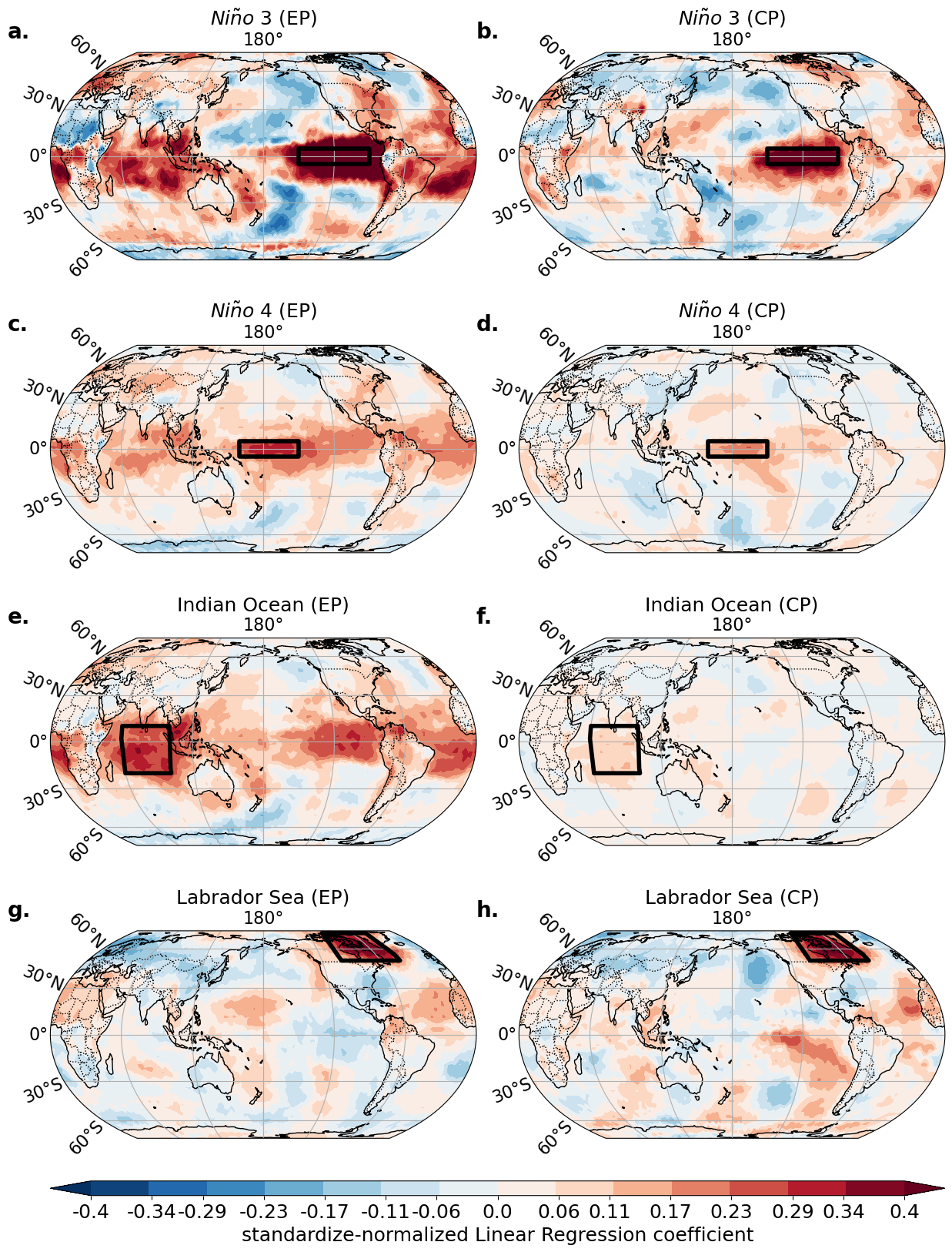}
    \caption{\textbf{SAT fields (Z-score normalized) regressed on temperatures from the four selected regions.}
    SAT time series at each location is regressed against all the rank-normalized SAT time series inside the four regions (black rectangles). Displayed are the Ni\~no 3 box (1. row), Ni\~no 4 box (2. row), Indian Ocean (3. row), and Labrador Sea (4. row), with first column denoting EP and second column CP conditions.}
    \label{fig:LR-coefficient_standardize}
\end{figure}

\begin{figure}[!tb]
    \centering
    \includegraphics[height=1.\textwidth]{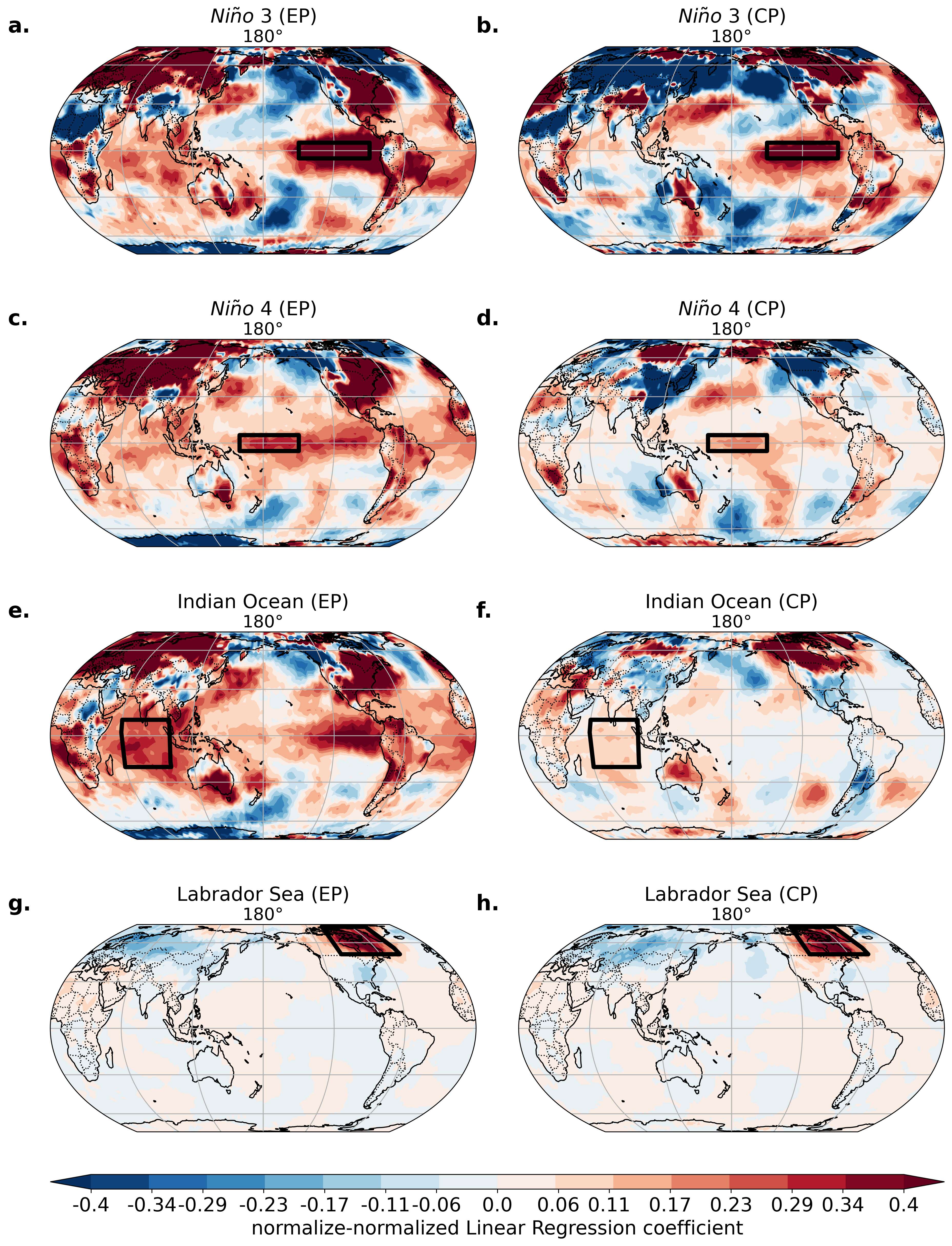}
    \caption{\textbf{SAT fields (min-max normalized) regressed on temperatures from the four selected regions.}
    SAT time series at each location is regressed against all the rank-normalized SAT time series inside the four regions (black rectangles). Displayed are the Ni\~no 3 box (1. row), Ni\~no 4 box (2. row), Indian Ocean (3. row), and Labrador Sea (4. row), with first column denoting EP and second column CP conditions.}
    \label{fig:LR-coefficient_0_1_normalized}
\end{figure}

\end{document}